\documentstyle[epsfig,floats,prb,aps]{revtex}

\begin{document}
\title{\bf $ $ \\
Andreev States in long shallow SNS constrictions}

\author{A. Lodder} 

\address{Faculteit Natuurkunde en Sterrenkunde, Vrije Universiteit,
         De Boelelaan 1081,\\ 1081 HV Amsterdam, The Netherlands}

\author{Yu. V. Nazarov}

\address{Faculteit der Technische Natuurkunde and DIMES, Technische Universiteit Delft,
         Lorentzweg 1,\\ 2628 CJ Delft, The Netherlands}

\date{\today}

\maketitle

\begin{abstract}
\normalsize{
We study Andreev bound states 
in a long shallow  normal constriction,
which is open to a superconductor  at both ends.
The interesting features of such setup include the absence 
of electron-hole symmetry and the interference of electron waves 
along the constriction.
We compare results of a numerical approach based on the Bogoliubov 
equations with those of a refined semiclassical description.
Three types of Andreev bound states occur in the constriction:
{\it i}) one where both electron and hole wave part 
of the bound state propagate through the 
constriction, {\it ii}) one where neither electron nor hole wave part
propagate, and {\it iii}) one where only the electron 
wave propagates.
We show that in a wide energy region the spacing between the Andreev
states is strongly modulated by the interference of electron waves
passing the constriction. 
}
\end{abstract}



\section{Introduction}

Andreev bound states \cite{Kulik} occur in a normal metal system of finite
size that is in contact with bulk superconductors. They exhibit
complicated and sometimes counter-intuitive physics which depends much
on disorder and electron scattering in the normal metal. The most adequate
method of describing Andreev states in metals is a semiclassical one. \cite{Larkin}
In many occasions the method can be reduced to a simple circuit theory. \cite{Nazarov} 

A new challenge is presented by the fabrication of systems where the normal part
is represented by a doped semiconductor. Typically, Andreev states are being
formed in a two-dimensional electron gas. There are means to shape such a 2DEG
in different ways forming controllabe constrictions. There are also means to 
provide conditions for ballistic transport of the electrons, so that they are scattered
at the constriction boundaries only. The Andreev states in such systems can also
be addressed by the traditional semiclassical technique, see e.g. \cite{YuliLod}
However there are cases where this approach fails.

The point is that the traditional semiclassical approach essentially exploits an
approximate electron-hole symmetry. This symmetry holds provided the
superconducting energy gap $\Delta$ exceeds a typical energy of the electrons. 
Recently it has been pointed out that this condition may fail in few-mode 
superconductor-semiconductor-superconductor constrictions. \cite{Chtchelkatchev}

\begin{figure}[htb]
\centerline{\epsfig{figure=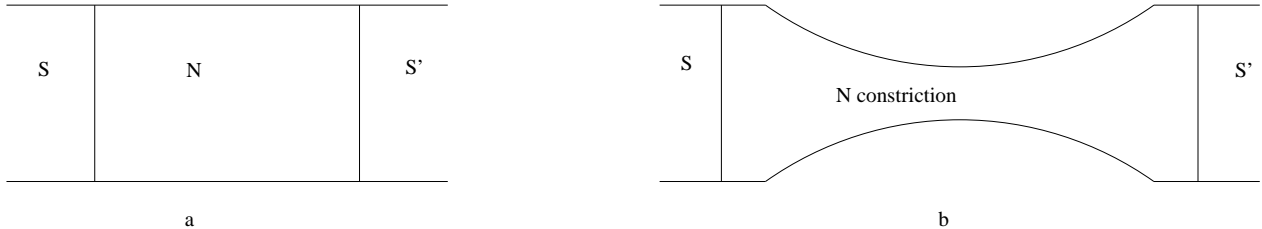,height=3.0cm}}
\caption[]{A straight SNS junction
and a SNS constriction.}
\label{SNSC}
\end{figure}

If electron-hole symmetry does not hold, the problem should be adressed by means
of the Bogoliubov equations that provide a strict quantummechanical treatment. 
The equations can be easily solved for a separable geometry.
That is why the occurrence and character of Andreev bound states in
a "straight" SNS junction are well understood.\cite{Miriam} A two-dimensional picture
of such a system is shown in FIG. \ref{SNSC}a.  The simplicity of the problem comes
from the simple shape of the boundaries. 

As soon as the boundaries are curved, the geometry is not separable any more.
The quantummechanical problem appears to become hardly solvable acquiring 
features of quantum chaos. 
In the present paper we present a simple model system where complications related
to chaos may be lifted but the absence of the electron-hole symmetry plays an important
role.  We have
chosen for a SNS junction with a constriction in the middle, schematically depicted
in FIG. \ref{SNSC}b. We assume the adiabaticity of the confining potential that
allows to reduce the problem to a set of one-dimensional problems for non-mixing
transverse modes.\cite{Glazman} We solve the one-dimensional
Bogoliubov equations numerically.
To understand the results, and the degree of their generality, we present in addition
two semiclassical methods for the one-dimensional problem.

The paper is organized as follows.
In Section \ref{theory} we derive the Bogoliubov equations for the adiabatic
constriction. We specify the model in use and analyze related energy 
scales in Sec. \ref{model}. We present a simple semiclassical
picture of Andreev levels in the Sec. \ref{Semiclassical}. A
brief qualitative description of the results is given 
in Sec. \ref{qualitative results}. The numerical results
are presented and discussed in Sec. \ref{results}.
We give concluding remarks in Sec. \ref{conclusions}.
An interpretation of the interference features found is 
given in Appendix \ref{quasiclasmod}, along with the underlying
semiclassical description.

\section{The Bogoliubov equations for the model constriction}
\label{theory}

The Bogoliubov equations for a clean system with free electrons
in the normal metallic part of the system can be written as
\begin{eqnarray}
  \label{Bogol3}
\left[\begin{array}{cc}
  -\frac{\hbar^2}{2m} {\nabla}^2 - \mu &   \Delta({\bf r}) \\
 \Delta^{\ast}({\bf r}) &
 \frac{\hbar^2}{2m}{\nabla}^2 + \mu 
\end{array}\right]
\Psi({\bf r}) = E \Psi({\bf r})\equiv E\left(\begin{array}{c}u({\bf r}) \\
 v({\bf r}) \end{array}  \right).
\end{eqnarray}
The two-component wave function $\Psi({\bf r})$ 
describes quasiparticle excitations
of the superconducting state,
the energy $E$ of those is being counted from the 
Fermi energy $\mu$.
The spatial dependence of the gap function $\Delta({\bf r})$
remindes that we deal with an inhomogeneous
system. In the superconductors  $\Delta({\bf r})$
is assumed to be a complex constant, while in the normal part
of the system
$\Delta({\bf r}) = 0$. We restrict ourselves to effectively
two-dimensional systems. This is achieved technically by
choosing the thickness $w$ in one direction, let us say the
$z$ direction, such that $\frac{\hbar^2}{2m}{(\frac{\pi}{w})}^2>\mu$,
by which transverse modes in the $z$ direction are forbidden.
 
In the following derivation we go along the path proposed 
in \cite{Glazman} for normal constrictions.
We assume that the boundary potential in the remaining transverse
direction, the y-direction, is infinitely high, so that the
wave function must be equal to zero at the transverse boundaries,
which can be expressed by a sine function. The total wave function
$\Psi$, which is now a function of two coordinates, then can
be presented as the linear combination of the following form
\begin{equation}
  \label{TwoDwave}
\Psi(x, y) = \sum_n \phi_n(x)
\sin\left(\frac{n\pi}{d(x)}(y - \frac{d(x)}{2})\right),
\end{equation}
$\phi_n$ being a two-component wavefunction.
Substituting this form in Eq. (\ref{Bogol3}) and
taking the inner product with $\sin\left(\frac{n\pi}{d(x)}
(y - \frac{d(x)}{2})\right)$, 
one obtains the equations for the upper component of $\phi_n$
\begin{eqnarray}
   \label{Coupled}
&\lbrack-{\phi}_n^{''}(x) + (\frac{n^2\pi^2}{d^2 (x)}-\frac{2 m\mu}{\hbar^2}){\phi}_n(x)
-\frac{d^{'}(x)}{d(x)}{\phi}_n^{'}(x) 
 +{\phi}_n(x)\lbrace{\left(\frac{d^{'}(x)}{d(x)}\right)}^2\left(\frac{n^2\pi^2}{12}
+\frac{1}{2}\right)- 
\frac{d^{''}(x)}{2d(x)} \rbrace\rbrack \nonumber \\
& -2{\sum}_{p\neq n}^{p\pm n\hspace{1mm}{\rm even}}\lbrack
p\frac{d^{'}(x)}{d(x)}{\phi}_p^{'}(x) \left(\frac{1}{n+p}
+\frac{1}{n-p}\right)+{\phi}_p(x)\lbrace{\left(\frac{pd^{'}(x)}{d(x)}\right)}^2
\left(\frac{1}{{(n+p)}^2}-\frac{1}{{(n-p)}^2}\right)-\nonumber \\
&p\left({\left(\frac{d^{'}(x)}{d(x)}\right)}^2 -\frac{d^{''}(x)}{2d(x)}\right)
 \left(\frac{1}{n+p} + \frac{1}{n-p}\right)\rbrace\rbrack =
\frac{2m}{\hbar^2} E{\phi}_n(x).
\end{eqnarray}
The corresponding equation for the lower component 
differs from Eq. (\ref{Coupled})
only by a minus sign of the left hand side.
The components are decoupled, since in the constriction 
the superconducting gap parameter $\Delta$ equals zero.
For the time being we will work with
a slowly varying transverse shape of the constriction, and
assume that the terms containing derivatives $d^{'}(x)$ and
$d^{''}(x)$ can be neglected. After we have specified the system we will give
an estimate of these terms. In this approximation, the 
different transverse modes  are uncoupled. Therefore we obtain
the following equation for each mode
\begin{eqnarray}
  \label{Bogol1}
\left[\begin{array}{cc}
   -\frac{\hbar^2}{2m}\frac{d^2}{{dx}^2} +U(x) - \mu &   \Delta(x) \\
 \Delta^{\ast}(x) &
 \frac{\hbar^2}{2m}\frac{d^2}{{dx}^2} -U(x) + \mu
\end{array}\right]
\phi_n(x) = E \phi_n(x)
\end{eqnarray}
where we have defined an effective potential
$U(x) \equiv {\frac{\hbar^2}{2m}(\frac{n\pi}{d(x)})}^2$,
representing the transverse mode energy.
The two-component wave function in the longitudinal 
direction of the
system, $\phi_n (x)$, assumes different forms in
the different parts of the system.
In the superconductor left of the normal metal it has the form
  \begin{eqnarray}
  \label{solSA}
 && A_+  \left(\begin{array}{c}
      e^{i\frac{\phi_{\rm L}}{2}}\sqrt{E-\sqrt{E^2-|\Delta_{\rm L}|^2}}
\\e^{-i\frac{\phi_{\rm L}}{2}}\sqrt{E+\sqrt{E^2-|\Delta_{\rm L}|^2}}
    \end{array}\right)e^{ix\sqrt{k_{Fmax}^2-\frac{2m}{\hbar^2}\sqrt{E^2-|\Delta_{\rm L}|^2}}} +
     \nonumber \\
 && A_-  \left(\begin{array}{c}
      e^{i\frac{\phi_{\rm L}}{2}}\sqrt{E+\sqrt{E^2-|\Delta_{\rm L}|^2}}
\\e^{-i\frac{\phi_{\rm L}}{2}}\sqrt{E-\sqrt{E^2-|\Delta_{\rm L}|^2}}
    \end{array}\right)e^{-ix\sqrt{k_{Fmax}^2+\frac{2m}{\hbar^2}\sqrt{E^2-|\Delta_{\rm L}|^2}}},
  \end{eqnarray}
where $\Delta_{\rm L}=e^{i\phi_{\rm L}}|\Delta_{\rm L}|$
and $k_{Fmax}^2 \equiv 2m\mu/\hbar^2 - \frac{n^2{\pi}^2}{d^{2}_{max}}$,
$d_{max}$ being the width in the superconductor.
Similarly, at the right hand side of the constriction this form 
becomes
  \begin{eqnarray}
  \label{solSB}
 && B_+  \left(\begin{array}{c}
      e^{i\frac{\phi_{\rm R}}{2}}\sqrt{E+\sqrt{E^2-|\Delta_{\rm R}|^2}}
\\e^{-i\frac{\phi_{\rm R}}{2}}\sqrt{E-\sqrt{E^2-|\Delta_{\rm R}|^2}}
    \end{array}\right)e^{ix\sqrt{k_{Fmax}^2+\frac{2m}{\hbar^2}\sqrt{E^2-|\Delta_{\rm R}|^2}}} +
     \nonumber \\
 && B_-  \left(\begin{array}{c}
      e^{i\frac{\phi_{\rm R}}{2}}\sqrt{E-\sqrt{E^2-|\Delta_{\rm R}|^2}}
\\e^{-i\frac{\phi_{\rm R}}{2}}\sqrt{E+\sqrt{E^2-|\Delta_{\rm R}|^2}}
    \end{array}\right)e^{-ix\sqrt{k_{Fmax}^2-\frac{2m}{\hbar^2}\sqrt{E^2-|\Delta_{\rm R}|^2}}},
  \end{eqnarray}
\begin{figure}[tb]
\centerline{\epsfig{figure=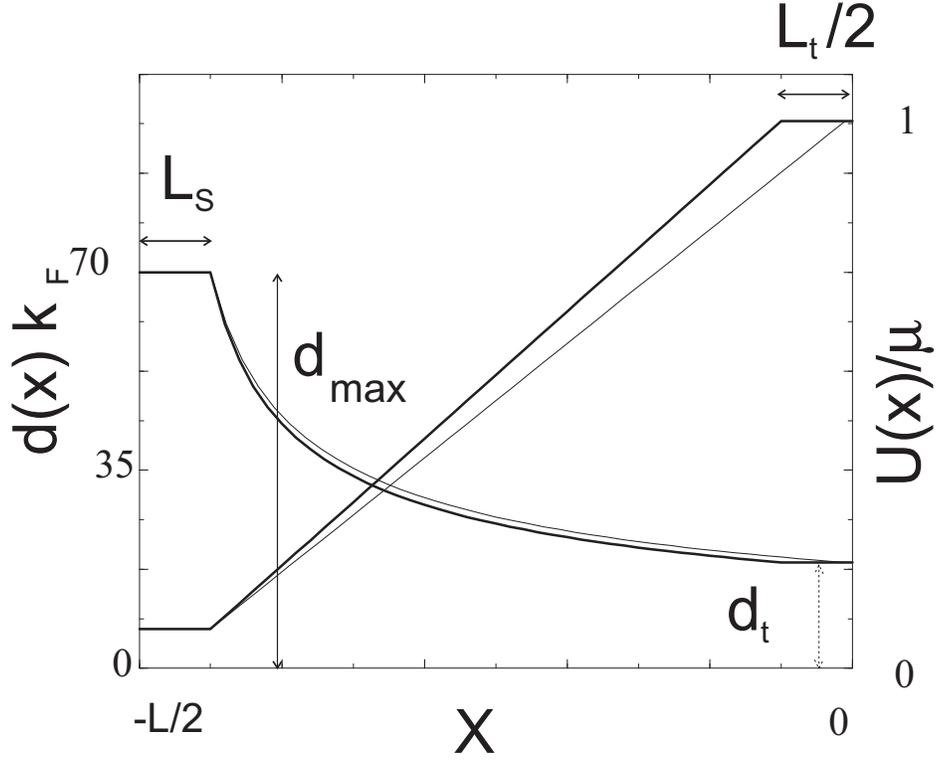,height=10.0cm}}
\caption[]{The shape of the model constriction
that corresponds to a linear potential U(x), for two different $L_t$.}
\label{system}
\end{figure}
Obviously, the labels L and R refer to the location of the superconducting
part with respect to the constriction.

The above forms shall be used as boundary conditions
for Eq. (\ref{Bogol1}).


\section{Model}
\label{model}
The model constriction is  depicted in
FIG. \ref{system}. The constriction is symmetric.
To simplify numerical calculations, we choose the shape of the boundaries in
such a way that the effective one-dimensional 
potential $U(x)$ either remains constant of varies linearily.
This choice has the advantage that the solutions of Eq. (\ref{Bogol1})
can be readily expressed in terms of well-known Airy functions. 
The effective potential and the corresponding shape of the constriction,
$d(x)$, are shown in FIG. \ref{system}.

The figure shows only half of the junction, the part in the
negative $x$ direction.
The constriction is {\it long} so that $L k_F \gg 1$. The constriction
is {\it shallow} so that the top of the potential $U_{max}$ for a
transverse mode of interest
almost matches
Fermi energy, $U_{max} \approx \mu$.
 We thus define $\delta E \equiv \mu - U_{max}$, $|\delta E| \ll \mu$.
This is achieved by a proper choice
of the width $d_t$ in the middle of the constriction. An important
parameter is also $L_t$, the length of the narrowest part of the
constriction. The width reaches its maximum value $d_{max}$ at
$|x|=(L -L_t)/2$, which is at a distance $L_s$ from the superconductors.

\begin{figure}[tb]
\centerline{\epsfig{figure=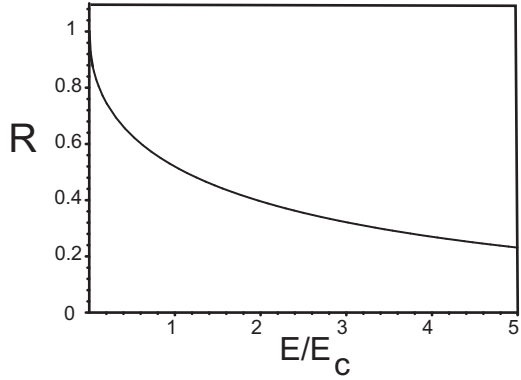, height=5cm}}
\caption[]{The reflection coefficient from the top as function of energy.}
\label{reflect}
\end{figure}

Despite apparent simplicity of the model, it exhibits a variety
of relevant energy scales. We list them going from bigger ones to smaller ones.
The biggest scale is obviously the Fermi energy itself, $\mu$. 
Another important scale we obtain by considering electron reflection
from one side of the top and disregarding existence of another side.
Classically, the electron is fully transmitted at energies exceeding $U_{max}$
and fully reflected otherwise. Due to quantum effects, there is a
smooth change from full reflection
to full transmission that occurs within an energy interval $E_c$. There is a related
length scale $x_c$ that corresponds to minimal electron wavelength. We
estimate the typical length scale $x_c$ and the energy scale $E_c$ by equating
potential and kinetic energy: $E_c \simeq (dU/dx) x_c \simeq \hbar^2/m x_c^2$.
Let us note that by virtue of our model the potential derivative can be estimated 
as $dU/dx \simeq \mu/L$. 
By that we define an energy scale $E_c = 0.03 \mu (k_F L)^{-2/3}$ at which
the reflection of electron waves from the top is changing to
full transmission. The numerical coefficient is choosen in such a way
that the reflection probability reaches $0.5$ at $E-U_{max} \approx E_c$.
The energy dependence of the reflection probability is plotted in FIG. \ref{reflect}.
The curve roughly behaves as $R\approx E_c/E$.

Three different energy scales arise from the interference of
the electron waves either
beyond or at the top of the potential. 
Beyond the top, the typical wavevector is
of the order of $k_F$ and the interference condition
reads $\delta k L \simeq \pi$. 
This gives $E_{i1} \simeq \hbar^2 k_F\delta k/m \approx \pi\mu {(k_F L)}^{-1}$.
The typical traversal
time for this part of the system is of the order of $\hbar/E_{i1}$.
The wavevector at the top is considerably smaller,
$k_{top}\approx {(m\delta E)}^{\frac{1}{2}}/\hbar$. This may lead to much longer
traversal times through the top of the constriction. Substituting
$\delta E \simeq E_c$ we obtain the corresponding energy scale
$E_{i2} = (2 L/L_t) (E_c/\mu)^{1/2} E_{i1}$, and $E_{i2} \ll E_{i1}$ if $L
\simeq L_t$.

The third energy scale we obtain by estimating
the energy counted from the top at which a few wavelengths match 
the length of the top, $k_{top} L_t \simeq 1$.
This yields $E_{i3} \simeq \mu (k_FL_t)^{-2}$.
The same energy scale $E_{i3}$ determines when tunneling through the top
becomes essential. We see that $E_{i2} \ll E_c$, at least if $L \simeq L_t$. 
Usual assumptions for adiabatic constrictions are such that $E_{i2} \simeq E_c$.\cite{Glazman}
This makes our theory distinct from, for instance, Ref. \cite{Chtchelkatchev}.

The above analysis of our model captures the essential features
of electron propagation in long shallow junctions and all qualitative results
are actually model-independent.

We have performed numerical calculations for constrictions
that have a length $L$ ranging from $10^5 k^{-1}_F$ to $10^6 k^{-1}_F$,
and  $d_{max}$ = $70 k^{-1}_F$ at the SN boundaries. The width
at the middle $d_t$ is chosen to
vary between 17 and 28 $k^{-1}_F$ that assures that 
$U_{max} \approx \mu$ for the transverse mode $n=6$. 

Using the Airy function solutions of Eq. (\ref{Bogol1}), we
estimate the errors involved in neglecting the terms
in Eq. (\ref{Coupled}), which contain derivatives of $d(x)$.
As far as the first derivative is concerned, looking at
the second and third term in Eq. (\ref{Coupled}), we
calculated the relative errors
\begin{equation}
  \label{derivative1}
\frac{d^{'}(x)Ai^{'}(x)}{(\mu - U(x))d(x)Ai(x)}\hspace{5mm}
{\rm and}\hspace{5mm}\frac{d^{'}(x)Bi^{'}(x)}{(\mu - U(x))d(x)Bi(x)}
\end{equation}
for the two Airy functions $Ai(x)$ and $Bi(x)$, for three  $x$ values,
at the beginning of the constriction, somewhere in the middle,
and at the position where the narrowing has come to an end.
Looking at the fourth term, and realizing that for our
linear potential $U(x)=tx+c$ the relation $d^{''}d=
3 {d^{'}}^2$ holds, we calculated the relative error
\begin{equation}
  \label{derivative2}
\frac{{\left(\frac{d^{'}(x)}{d(x)}\right)}^2\left(\frac{n^2\pi^2}{12}
+\frac{1}{2}\right)- 
\frac{d^{''}(x)}{2d(x)}}{\mu - U(x)}=
\frac{\left(\frac{n^2{\pi}^2}{12} - 1\right)
{\left(\frac{d^{'}(x)}{d(x)}\right)}^2} {\mu - U(x)}
\end{equation}
at the same positions as well. For the three points mentioned
and for five different energies smaller than $|\Delta|$ the
average error in the first derivative appeared to be
smaller than $0.01\%$, and the error in the second
derivative was even a factor of $10^3$ smaller. It could
be concluded, that the omitted terms are really negligible.

\section{Semiclassical Andreev states}
\label{Semiclassical}

Let us start by presenting a semiclassical picture of Andreev states
in the simplest case where electrons and holes are either fully transmitted
or fully reflected from the potential $U(x)$.  
We count energies of the states from the chemical potential. 
We denote $\delta E \equiv \mu - U_{max}$.

\begin{figure}[tb]
\centerline{\epsfig{figure=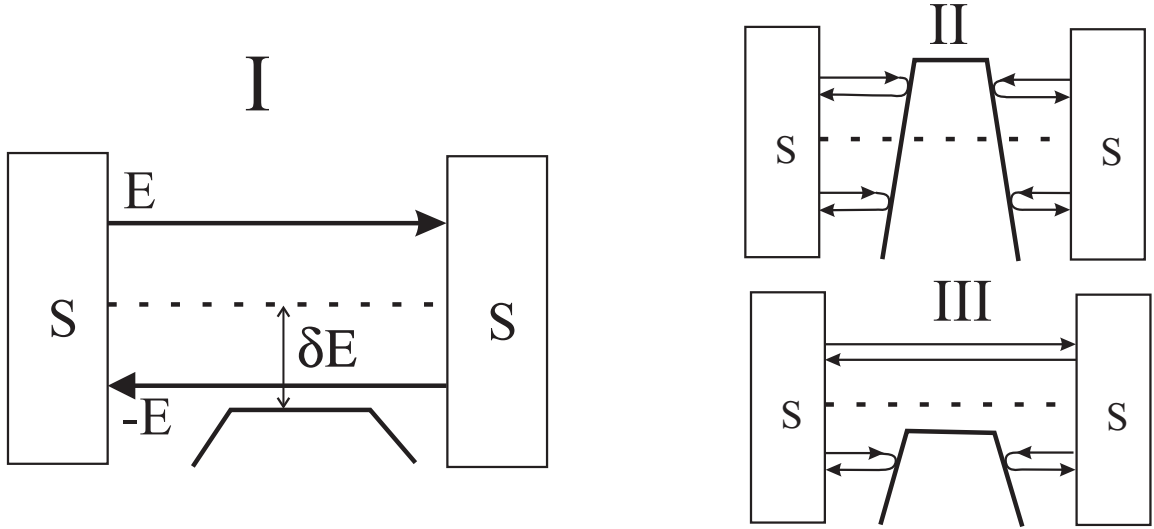,height=7.0cm}}
\caption[]{Three types of semiclassical Andreev states. The type III is specific for
the absence of electron-hole symmetry.}
\label{thereandback}
\end{figure}

We assume no phase difference between superconductors.
The Andreev state with positive energy $E$ is made of electron waves
with energies $E$ and $-E$, those we call "electrons" and "holes" respectively.
There may be three situations, depicted in FIG.\ref{thereandback}:
I. Both electron and hole are fully transmitted through the constriction,
$\delta E >0$ and $E<\delta E$. 
II. Both electron and hole are reflected from the top, $\delta E<0$ and
$E<|\delta E|$.
III. The electron is transmitted and the hole is reflected, $E>|\delta E|$.
Since the potential $U(x)$ varies slowly, 
the quantized energies of Andreev states may be obtained from the Bohr-Sommerfeld
quantization rule applied to Eq. (\ref{Bogol1}).
 The solutions of Eq. (\ref{Bogol1}) can be written in the form
\begin{equation}
  \label{ETrelation}
e^{i\int^x k_+(s)ds} \hspace{2mm}{\rm and}\hspace{2mm}e^{i\int^x k_-(s)ds}
\end{equation} 
for the electron and hole part of $\phi_n(x)$ respectively, where
$k_{\pm} = \sqrt{2m(\mu - U(x) \pm E)}/\hbar$.
We begin our discussion  with the case I.
In this case, an electron travelling from left to
the right builds up a phase of $\int_{-L/2}^{L/2} k_+(x)dx$. The electron
undergoes Andreev reflection and returns as a hole, and the wave function gains
the additional phase $\phi_A = \rm{acos} (E/\Delta)$, see Eqs. (\ref{solSA})
and (\ref{solSB}). The hole builds up a phase of
$\int_{L/2}^{-L/2} k_-(x)dx$. After Andreev reflection at the left side, the particle
has made a complete roundtrip. The quantization condition is that the phase gain is a
mupliple of $2 \pi$, that is
\begin{equation} 
  \label{ETrelation2}
\int_{-L/2}^{L/2} (k_+(x)-k_-(x))dx +2 \phi_A = 2\pi n 
\label{quantization_condition}
\end{equation}
Differentiating Eq. (\ref{quantization_condition}) with respect to $E$ gives an 
approximate
relation for the energy difference between adjacent Andreev states,
\begin{equation}
\label{spacing}
E_{n+1}-E_{n} = {2 \pi \hbar} ( T_{el} + T_{hole} + 
2 \hbar \frac{\partial \phi_A}{\partial E})^{-1}
\end{equation}
where we have introduced semiclassical times of travelling from
the one interface to the other one, for electrons and holes respectively,
\begin{equation}
\label{Telhole}
T_{\rm el}=\int_{-L/2}^{L/2}\frac{dx}{\sqrt{2(\mu - U(x) + E)/m}}
\hspace{2mm}{\rm and}\hspace{2mm}T_{\rm hole} =\int_{-L/2}^{L/2}\frac{dx}
{\sqrt{2(\mu - U(x) - E)/m}}.
\end{equation}
and made use of semiclassical relation between the travelling
time and the derivative of the quantummechanical phase with respect to the energy.
If there are  $many$ bound Andreev states, their typical spacing shall 
be small being compared to $\Delta$, so that $T_{el,hole} \gg \hbar/\Delta$.
Since the derivative of the Andreev phase is of the order of $1/\Delta$, the 
second term in the right hand side of Eq. (\ref{spacing}) can be  disregarded
and $E_{n+1}-E_n \approx 2\pi\hbar{(T_{el}+T_{hole})}^{-1}$
This suggests a presentation of  our numerical results:
we will plot the energy spacing times the travelling time as a
function of the state energy.
 For the sake of symmetry we will use
\begin{equation} 
  \label{ETav}
\frac{1}{2 \pi \hbar}(E_{n+1}-E_n)\hspace{1mm}{\rm Av}\lbrace T_{\rm el}
+T_{\rm hole}\rbrace \equiv \Theta,
\end{equation}
where ${\rm Av}\lbrace T_{\rm el}+T_{\rm hole}\rbrace = \frac{1}{2}
(T_{\rm el}(E_n)+T_{\rm hole}(E_n)+T_{\rm el}(E_{n+1})+T_{\rm hole}(E_{n+1}))$.
The right hand side just defines the compact notation $\Theta$, and
$\Theta \rightarrow 1$
in the classical limit. It is important to note
that the Andreev states in this situation are degenerate. The 
state with electrons going to the right and holes going
to the left has precisely the same energy as the state with
electrons going to the left and holes going to the right.  
To be closer to our concrete model, let us introduce $T_b$, the
travel time from one of the interfaces to the top, and $T_t$,
the travel time through the top. The full traversal time for the case I is
then $4 T_b + 2 T_t$.

In case II the electron coming from the left is reflected from the top,
gets back to the left interface, undergoes Andreev reflection, 
gets to the right as
a hole, gets back and finally undergoes yet another Andreev reflection.
The energy spacing is again given by the 
inverse of the time of semiclassical motion.
This time is now $4 T_b$. The Andreev states on the left
side of the constriction do not overlap with the ones on
the right side. This also leads to two-fold degeneracy
of Andreev states for our symmetric setup.

In the case III the electron coming from the left passes
the constriction and undergoes Andreev reflection on the right side.
The resulting hole goes to the left, is reflected from the top
and undergoes Andreev reflection on the right side again.
This results in an electron going to the left that undergoes
Andreev reflection on the left side. There are thus four Andreev
reflections involved. The state encompasses the electron and
hole waves that propagate in both directions. Therefore it is not
degenerate. The travelling time is now $8 T_b + 2 T_t$.

If $T_t \ll T_b$, the spacing of the states of the types I and II
are evidently the same. For type III states, the travelling time
is now approximately 
two times longer, which corresponds to a two times lesser energy
spacing. However, there is no degeneracy in this case.
This is why the number of states per energy interval
remains approximately the same in all three cases.
If $T_t \gg T_b$, there is a big spacing between type II states
corresponding to energy scale $E_{i1}$. The spacing of the
type I and type II states is much smaller.

Even a short glance at our numerical results
proves that the picture based on the Eqs. (\ref{ETrelation})
and (\ref{ETrelation2}) is oversimplified. In the following section
we present a more sophisticated picture, which accounts for
the oscillating phenomena,
found from the full solution of the
Bogoliubov equations (\ref{Bogol1}).


\section{Qualitative results}
\label{qualitative results}
\begin{figure}[tb]
\centerline{\epsfig{figure=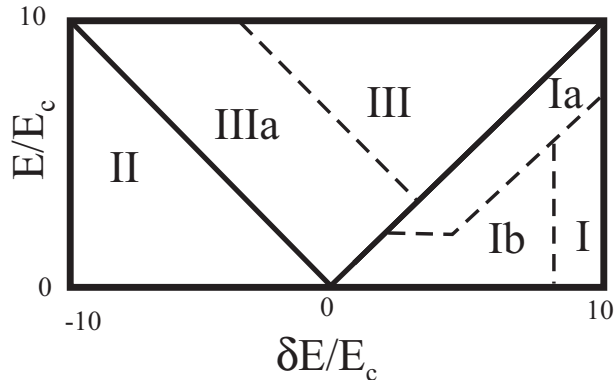, height=5cm}}
\caption[]{The parameter regions were different types
of Andreev states occur.
The reflection from the top causes interference that gives rise to
electron, hole and irregular oscillations in the regions IIIa, Ia and Ib respectively.}
\label{threeregions}
\end{figure}

Now we are in the position to present a concise summary of our qualitative results. 
Relevant parameters that determine the behaviour of the Andreev states are their 
energy $E$ and the shift $\delta E$ of the chemical potential with respect to
the potential maximum. The superconducting gap $\Delta$ just determines
the energy interval where Andreev states may be present, provided there are
many states in this interval.

The parameter regions exhibiting distinct behavior of Andreev
states are presented in FIG. \ref{threeregions}. Two diagonal lines correspond to either
electron or hole part of the state crossing the top of the potential
relief. These lines separate the
regions I, II and III discussed in the previous section. In these
regions, the Andreev states are almost equally spaced, with the spacing
being a slow function of energy. The energy scale of the spacing
corresponds to $ {\it min}(E_{i2},E_{i1})$. The semiclassical theory relates
the spacing with the classical traveling time.
From this one would conjecture a sharp crossover between the regions
I, II, III, this presumably  taking  place in an energy interval of the order
of several level spacings. This in fact corresponds to the situation
described in Ref. \cite{Chtchelkatchev}. 

The transitions between the regions I, II and III are much less
trivial for a long shallow constriction described here.
The reason for that is a strong reflection from the top of
the constriction that takes place in a wide energy interval $\simeq E_c$
near the lines, see FIG. \ref{reflect}.  The reflection takes
place at two edges of the top. The interference between the reflected
and transmitted waves disrupts a simple picture of semiclassical
Andreev states. 

This is why we have three more regions, IIIa, Ia, and Ib in the diagram.
They are separated by dashed lines from their parent regions.
Whereas the solid lines denote sharp transitions between the regions,
the dashed lines correspond to smooth crossovers. For our concrete
model, the crossovers are really very blurred, and the extra regions
are difficult to separate from the parents. The reason for that
is a slow decrease of the reflection coefficient with the energy, $R \approx E_c/E$.
The crossovers may become sharper in a different model of a long
shallow constriction. The regions would remain the same.

It is relatively easy to comprehend the situation in the region
IIIa. There, the holes are completely reflected from the top and
the transmitted and reflected waves do not interfere. 
They do interfere for electron components. This results in a regular
modulation of spacings between Andreev states, which we call
{\it electron oscillations}.
We present in an Appendix a detailed theory of this effect. Here
we give a final relation for "energy spacing times travelling time" $\Theta(E)$,
see Eq. (\ref{Theta_osc}):
\begin{equation}
\label{Theta_osc_0}
\Theta(E) = 1 +\frac {2 T_b}{T_t+2 T_b}\frac{\cos(2 \phi_b)\sqrt{R} - R}
{1+R-2\cos(2 \phi_b)\sqrt{R}},
\end{equation} 
$R$ being the energy-dependent reflection coefficient from the edge
of the constriction middle  and $\phi_b$ being a linear function
of energy, $\phi_b(E)= {\rm const.} + \pi E/E_{i1}$. If $E_{i2} \ll E_{i1}$,
each electron oscillation encompasses many
Andreev states, of the order of $E_{i1}/E_{i2}$.

In the region Ia the situation is almost the opposite. Here, those
are holes that are strongly reflected from the top. The electrons
go at higher energy and are being transmitted much better. 
So one expects interference of holes only. We will call this
phenomenon {\it hole oscillations}. Due to the
slow decrease of the reflection coefficient mentioned above,
the remaining interference of electrons is substantial. Therefore,
the hole oscillations are not as regular as the electron ones. Also,
the oscillation period is set by the inverse of the time for electrons to
travel through the top, $T^{el}_t$. The ratio of $T^{h}_t/T^{el}_t$
is not parametrically big, so that the oscillation encompasses 
only few Andreev states. 
In our numerical results, we see alternating spacing of Andreev states
in the region Ia.

In the region Ib both electrons and holes exhibit reflections,
and the interference picture thus becomes complicated and at best 
quasi-periodic.
We will call this {\it irregular} oscillations.
The visible quasi-period is $E_{i1}$ again. It arises as a result of
interference of two reflections, one for the electron component
and another for the hole one.  These reflections are coupled by an Andreev
process. One would not see such period in the interference picture
of a normal electron.      

All these three types of oscillations are seen in our numerical results.
Another effect of the reflection
is that the degeneracy between Andreev states is lifted.
For our symmetric setup, the Andreev states can be either even or
odd, that is, symmetric or antisymmetric with respect to $x \rightarrow
-x$.

\section{Numerical results}
\label{results}

For numerics, we choose the transverse dimensions
in such a way that only six
modes can pass the constriction.
For the sixth mode, the Fermi level matches the potential
in the middle of the constriction and $|\delta E |\ll \mu$. 
We note that from now on $\Delta$ is chosen to be a real constant,
since we assume no phase difference between the superconductors.
If $\delta E =0$ only the electron can pass the constriction. 
As soon as
$\delta E<-\Delta$ no transmission occurs for the waves forming
Andreev states. Since the interesting
phenomena occur for the sixth mode and
for not too big values of $\delta E/\Delta$, we will show results
for $\delta E/\Delta$ taking values 10, 2, 1, 0.5, 0, -0.5, -1 and -2 only.

Most calculations were done using a gap value of $4 \times 10^{-5} \mu$,
which is sufficiently small to achieve the semiclassical limit under investigation.

\begin{figure}[htb]
\centerline{\epsfig{figure=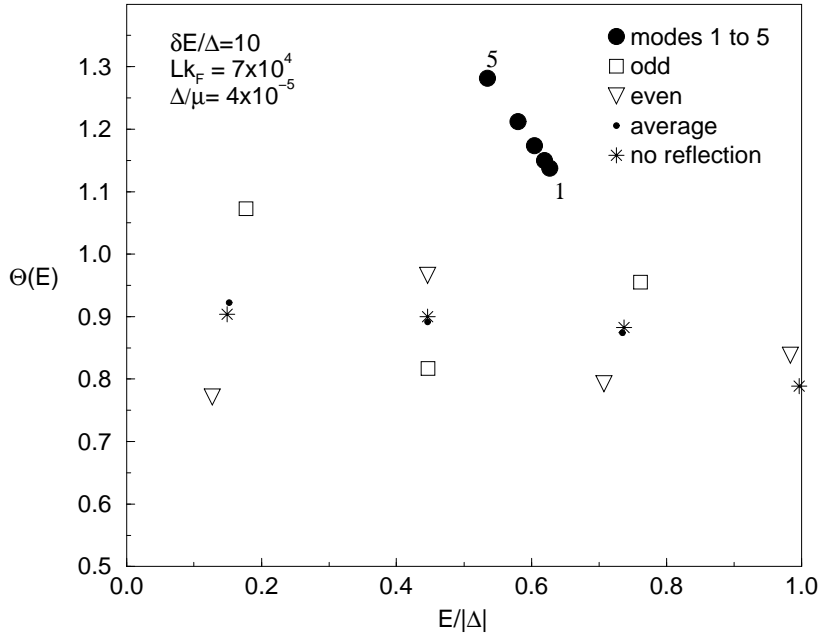,height=10.0cm}}
\caption[]{ $\Theta$ as defined by Eq. (\ref{ETav}) versus
the relative Andreev bound state energy $E/\Delta$, for a system length
$Lk_F= 7 \times 10^4$.
The first five modes support just one state each,
while for the sixth mode the number of states is seven.
Both electrons and holes traverse the system.}
\label{f10L1D2}
\end{figure}

\begin{figure}[htb]
\centerline{\epsfig{figure=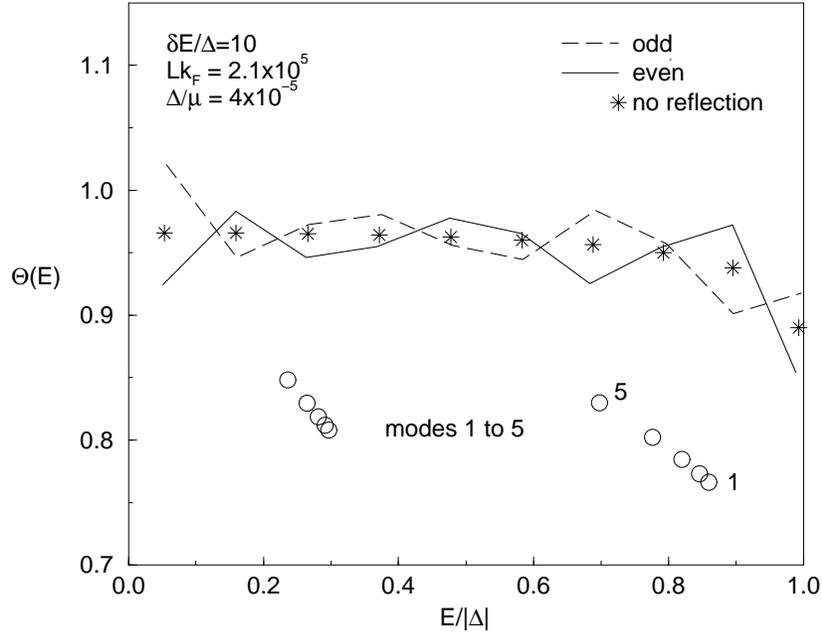,height=10.0cm}}
\caption[]{
The same as in FIG. \protect{\ref{f10L1D2}}, but for a three times longer system length
$Lk_F= 2.1 \times 10^5$. The reflection has increased resulting in bigger amplitude
of irregular oscillations of Andreev spacing. For the first five modes
two states are found
for each mode, while for the sixth mode the number of states is twenty.}
\label{f10L3D2}
\end{figure}

In order to get a feeling for the system properties we first show
the Andreev states spectrum for all modes and the two system lengths 
$L = 7 \times 10^4 k_F^{-1}$
and $2.1 \times 10^5 k_F^{-1}$ in the FIGs. \ref{f10L1D2} and \ref{f10L3D2} respectively.
For both cases $\delta E =10 \Delta$, so both components
of an Andreev state can transit the system.
The system lengths are such that the 
first five modes for $L = 7 \times 10^4 k_F^{-1}$
support one bound state each, while for the three times
longer system two bound bound states are found. The energies of these
states are plotted  with the circles.


The energy separation for the lower modes is approximately
$E_{i1}$, which is of the order of $\Delta$ choosen.
The wavefunctions of corresponding states easily pass the middle
part of the constriction and their energies do not depend on $\delta E$,
being very large for these states. Because they do not exhibit any interesting
behavior we do not pay further attention to
the first five modes. We just  mention, that in all results
to be displayed the Andreev approximation is made at the two SN
interfaces. In the exact treatment the degeneracy of the two
Andreev states with the two opposite propagation directions
is lifted, and one ends up with two standing waves, a symmetric
and an antisymmetric one. We calculated the exact energies for
the five states depicted in FIG. \ref{f10L1D2} by filled circles,
and found a splitting of the order of $10^{-10} \mu$, which is
negligible indeed.

More states are found for the sixth mode. Since they traverse
the top, their energy separation is determined by $E_{i2} \gg E_{i1}$
rather than $E_{i1}$.
We begin the discussion with FIG. \ref{f10L3D2}.
There are 
 20 Andreev states, as can be seen in the figure. Both
the odd and even states are shown, the corresponding
$\Theta$ values being connected by broken and solid lines respectively.
We see that the $\Theta$ values are close to 1, which proves
the validity of the semiclassical treatment. So we concentrate
on the deviations from 1.
The energy distance from the top of the potential
$\delta E \approx 47 E_c$.
Although the waves
are still well transmitted, the reflection probability at the narrow part
of the constriction becomes already appreciable (2\%).

This leads to two visible phenomena.
First, the two states corresponding to
two travelling waves propagating in opposite directions
are mixed by reflection. By that the even and odd states have different energies. 
The stars in the figure indicate
the result for an average energy, which
corresponds to two degenerate states found after neglect of
reflection at the constriction. The actual calculation in which only 
propagation to the right is allowed can be achieved by 
omitting the contribution of the wave propagating to the left in 
the normal metallic part at the left SN interface. The
corresponding $\Theta$ values are somewhat smaller than 1,
but they lie at a stable height, in agreement with the estimate
based on the forms (\ref{ETrelation}),
in which any reflection is not accounted for. The deviation 
of the stars from the lines is the measure of the effect.

Second, the interference between the reflected waves leads
to irregular dependence of $\Theta$ on energy, as discussed
in section \ref{qualitative results}. The amplitude of these
irregular oscillations is small corresponding to the small reflection
coefficient.

Both effects are better visible in FIG. \ref{f10L1D2} where the constriction
is shorter and only seven states are seen. They are
displayed by four triangles and three squares.
However, this is not the main difference between the figures.
The energy distance $\delta E$ becomes $\approx 23 E_c$, which
corresponds to a reflection coefficient of the order of 4\%.
The energy difference between the odd and even states
becomes considerable, even such, that the highest state, a symmetric
solution, does no longer have its antisymmetric counterpart.
This would have got an energy $E>\Delta$, while bound states
are only found below the gap. The irregular oscillations
are increased in amplitude.

\begin{figure}[htb]
\centerline{\epsfig{figure=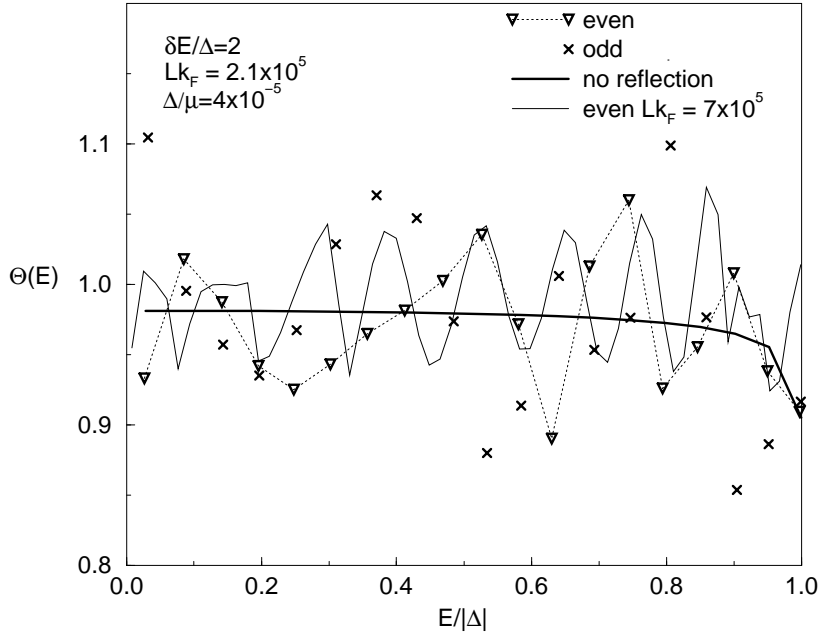,height=10.0cm}}
\caption[]{
The same as in FIG. \ref{f10L1D2}, but only for
mode 6 states. Irregular oscillations in the region Ib. The hole
branch of the Andreev states lies always an energy higher than the
gap energy above the barrier in the middle.
We show results for two system lengths $2.1\times 10^5$ and $7 \times 10^5$,
which give 38 and 119 bound states respectively. The reflection coefficient
varies between 0.05 and 0.1. The typical quaisiperiod is of the order
of $E_{i1}$. }
\label{f2D2}
\end{figure}
\begin{figure}[htb]
\centerline{\epsfig{figure=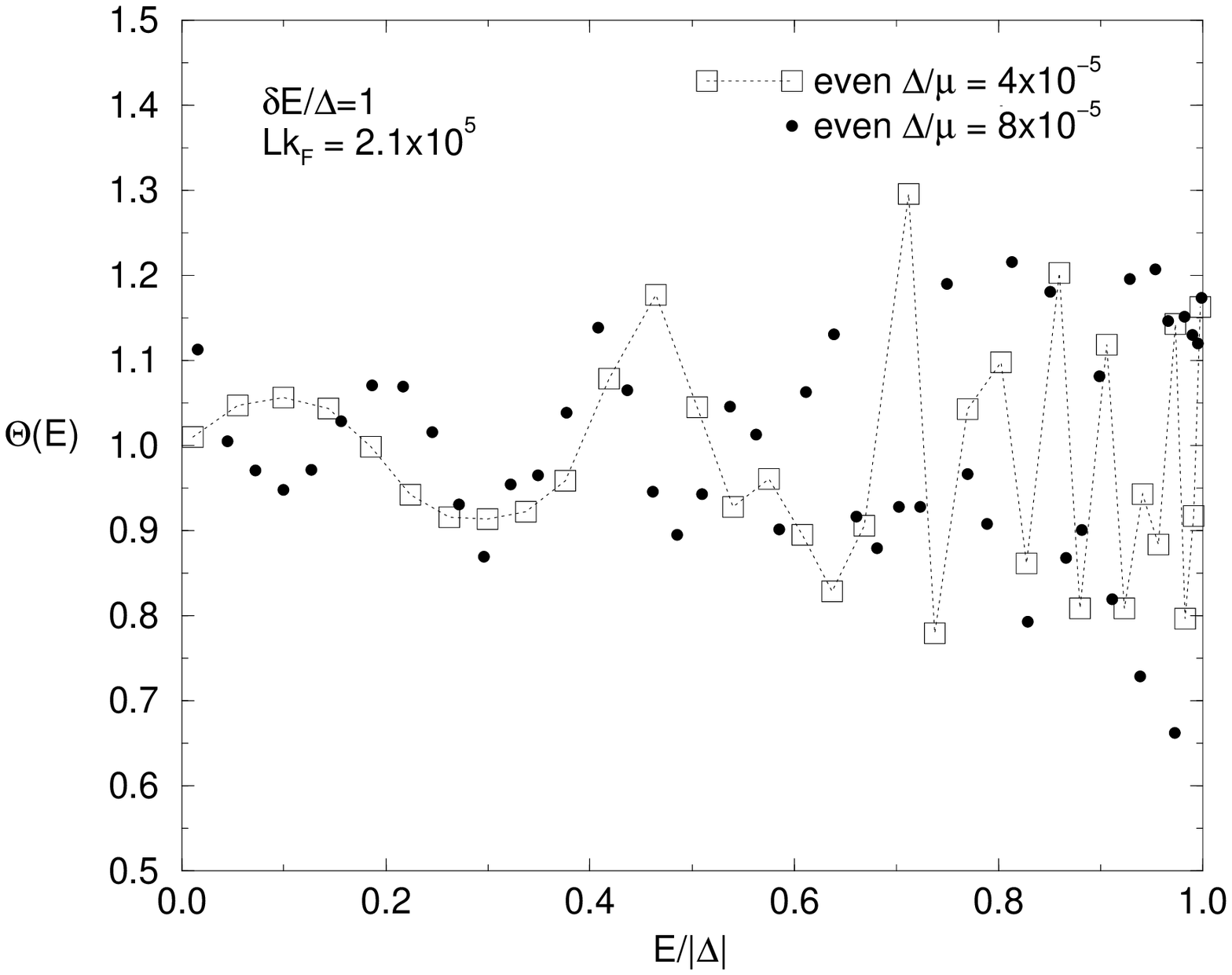,height=10.0cm}}
\caption[]{For $\delta E = \Delta$ the bound states near $\Delta$
have a hole component touching the top of the potential.
Strong reflection of these holes give rise to hole oscillations
above $\approx 0.7 \Delta$.
Two gap values correspond to 66 and 95 Andreev states.}
\label{f1L3}
\end{figure}

Results for $\delta E = 2 \Delta$ are shown in FIG. \ref{f2D2}.
In addition to the results for
the system length $L=2.1 \times 10^5$ the even states are shown for the
larger length of $L=7 \times 10^5$. The number of states has increased
to 119, and the oscillation frequency has increased as well.
The two choices of $L$ correspond to $\delta E/E_c \approx 10$ and
 $\delta E/E_c \approx 21$. The irregular oscillations are
bigger in the first case, which corresponds to the bigger
reflection. Due to larger number of states, one can see
already a typical (quasi)period. The biggest
interference scale $E_{i1}$ is $0.37 \Delta$ and $0.11 \Delta$
respectively. We see a good correspondence between this
scale and the quasiperiod. 
 
The next figure, FIG. \ref{f1L3}, presents the results for $ \delta E = \Delta$,
when Andreev states extend till the energy at which holes
are reflected completely. We observe a new type of oscillations
at energies $>0.6 \Delta$. At these energies, the holes are
reflected more effectively than electrons. So that the oscillations
are determined mostly by their interference. The two choices
of $\Delta$ correspond to $\delta E/E_c = 4.75$ and  $\delta E/E_c = 9.5$.

\begin{figure}[htb]
\centerline{\epsfig{figure=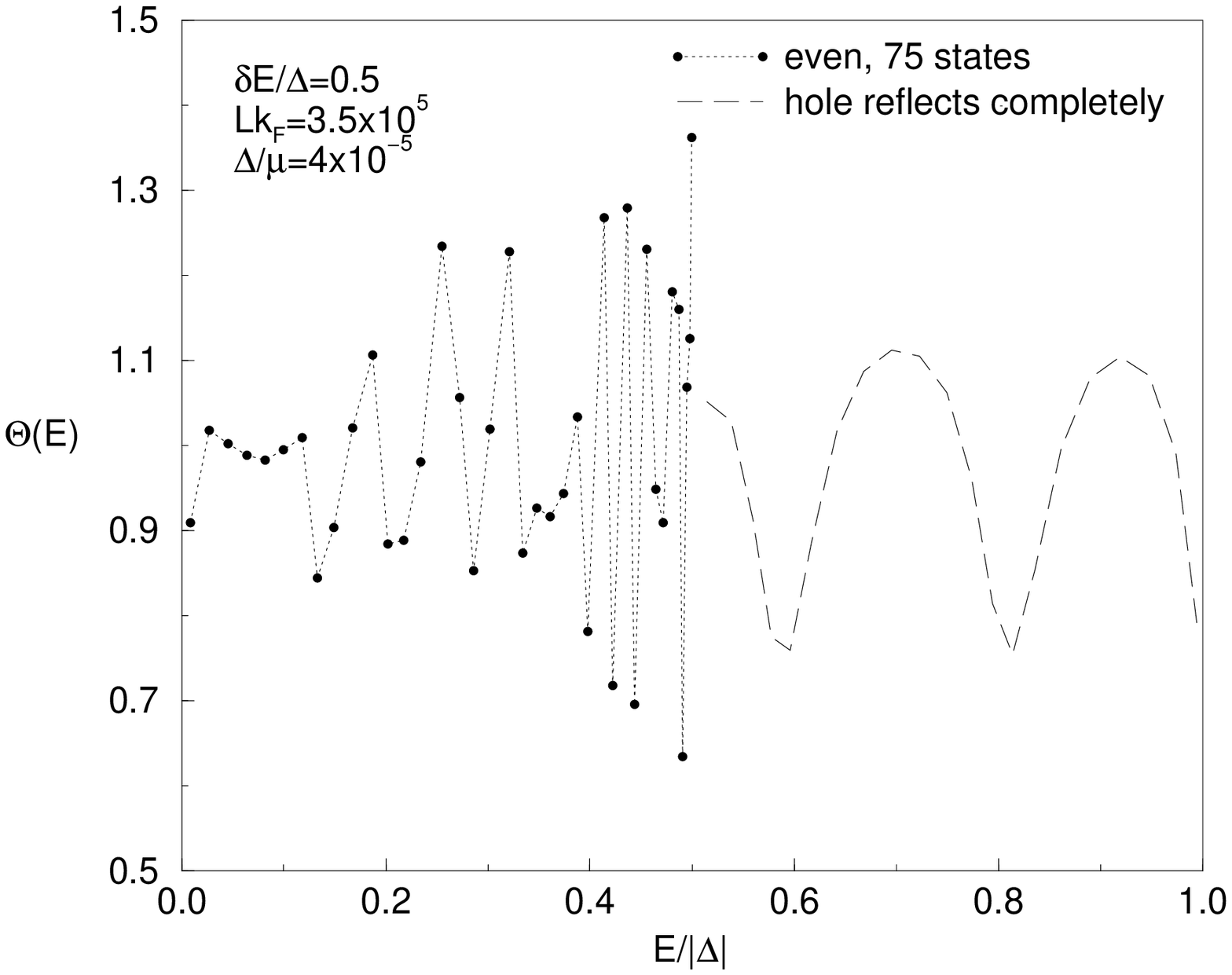,height=10.0cm}}
\caption[]{At $\delta E = 0.5 \Delta$ the hole component
is completely
reflected in the middle of the constriction at $E>0.5\Delta$.
All three types of oscillations are seen simultaneously in this plot.
}
\label{fphL5}
\end{figure}

The most interesting figure 
is probably  FIG. \ref{fphL5}, which
displays $\delta E  = 0.5 \Delta$ results. The parameter choice
corresponds to $\delta E \approx 3.3 E_c$. All three types
of oscillations can be simultaneously seen there.
The oscillations are irregular till $E \approx 0.3 \Delta$ and
they cross over to hole oscillations at higher energy. 
At $E=0.5 \Delta$ the holes are completely reflected,
and the interference picture is determined solely by electrons.
The transition is sharp indeed, the width being of the order 
$E_{i3}= 2 \times 10^{-5} \Delta$.
These electron oscillations are very regular.
Their period is determined by $E_{i1}= 0.22 \Delta$, 
each oscillation
encompassing about 10 Andreev states.
There is a striking contrast between the
electron and hole oscillations, the latter being
much less regular. 
The point is that the hole oscillations are disrupted 
by residual interference of electrons and the electron oscillations
are not, since  holes do not penetrate the middle of the 
constriction and do not interfere.
  
\begin{figure}[htb]
\centerline{\epsfig{figure=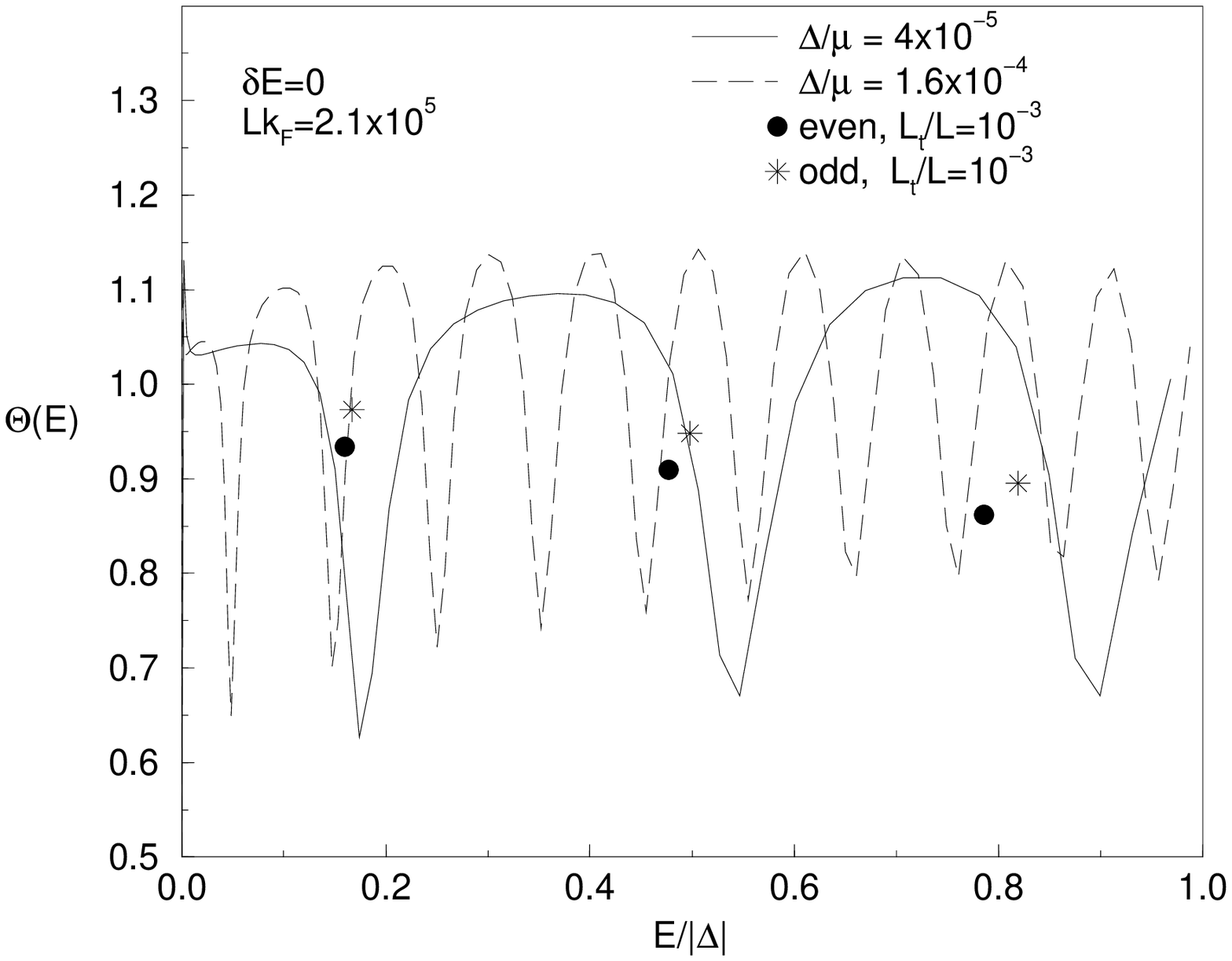,height=10.0cm}}
\caption[]{The electron oscillations through the whole
energy range. The hole component is completely reflected
 for all energies. Each oscillation encompasses about ten states.
 Circles and stars represent the states for a very short middle
 of the constriction and for the smaller gap value
of $\Delta = 4\times 10^{-5}\mu$. Note the correspondence of their positions
 and the oscillation dips.}
\label{f0L3}
\end{figure}

The electron oscillations throughout the full range of Andreev states
are presented in FIG. \ref{f0L3}. Here
the width of the constriction is chosen in such a way, that
$U(x)$ in the middle coincides with the Fermi energy and $\delta E =0$.
All
hole waves are reflected completely from the middle
of the constriction, while all electron waves still can pass.
The parameters correspond to $\Delta/E_c \approx 4.5$ for the solid curve
and $\Delta/E_c \approx 19$ for the dashed one. This
means that the reflection coefficient from the top varies from $1$ 
to $\approx 0.2$ or $\approx 0.05$ respectively. This affects
the oscillation amplitude.
 In FIG. \ref{f0L3} one indeed sees that the position of the
dips moves upwards with the energy, most clearly for the larger gap
value.
The spacing between the states and the oscillation period are determined
by energy scales $E_{i2}$ and $E_{i1}$ respectively. This is why for
four times larger gap four times more states and oscillations are found.

We remind that up to now all calculations were done with the length of 
a flat part in the middle
equal to $10\%$ of the system length, $L_t=0.1\ L$. 
We also show in FIG. \ref{f0L3}
the results for much shorter flat part length $ L_t=10^{-3} L$. 
Six states are found, three odd states and
three even states. For these parameters, the constriction middle 
is so short that $E_{i2} \gg E_{i1}$, and the spacing is of the order of
$E_{i1} \approx 0.37 \Delta$.  This is why the Andreev {\it states}
for the short top match the {\it oscillations} of Andreev states
for the long top. The matching is not perfect. The reason for that
is the tunneling of the holes through the top, which becomes 
considerable for the short top ($E_{i3} = 0.5 \Delta$). 
\begin{figure}[htb]
\centerline{\epsfig{figure=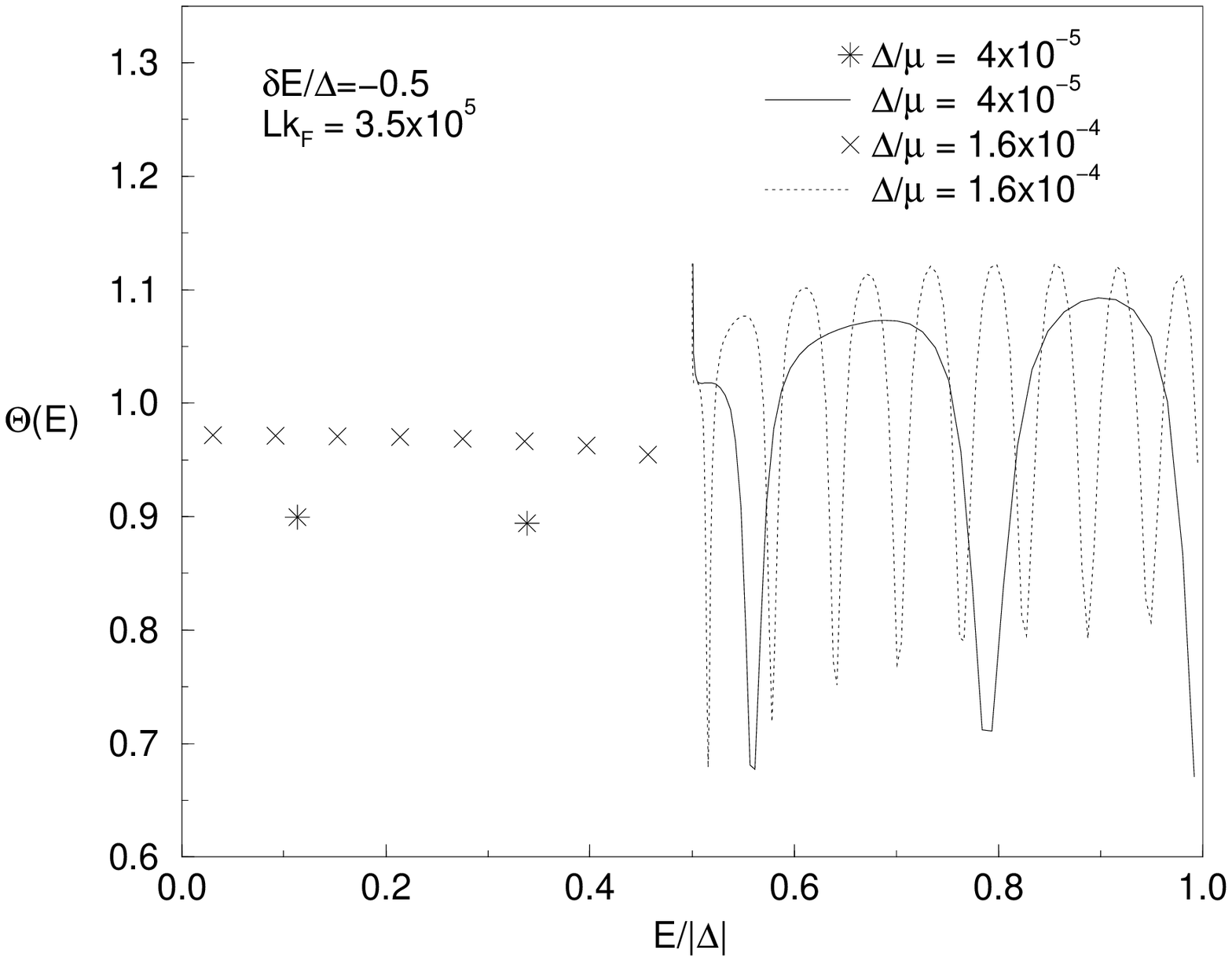,height=10.0cm}}
\caption[]{In this case, $E=0.5 \Delta$ separates regions II and III b.
The Andreev states in the region II are localized near the superconductors
and do not go through the middle of the constriction. Note the correspondence
between the spacing of the states at low energy and oscillation period
at higher energy.}
\label{fmhL5}
\end{figure} 

The following results are for parameter range $\delta E <0$
where holes are completely reflected from the top.
FIG. \ref{fmhL5}, with $\delta E = -\frac{1}{2}\Delta$,
is a counterpart of FIG.\ref{fphL5},
with  $\delta E = \frac{1}{2}\Delta$. 
This means that 
the electrons are completely reflected at $E<\frac{1}{2}\Delta$
and transmitted at higher energies.  By that the oscillations
arise for the higher energies only, which is clearly seen in the figure.
For $E<\frac{1}{2}\Delta$ the two parts of the constriction are
in fact not connected. The Andreev states
calculated are doubly degenerate, with one state resting in the left part
of the constriction and the other state in the right part. 
The spacing between these states is of the order of
$E_{i1}$ for both values of $\Delta$. Again, this
scale sets the oscillation period for the states above $\Delta/2$, whereas
their spacing is determined by much smaller energy scale $E_{i2}$.
Two values of $\Delta$ correspond to $\Delta/E_c \approx 4.5$ and
$\Delta/E_c \approx 19$.
\begin{figure}[htb]
\centerline{\epsfig{figure=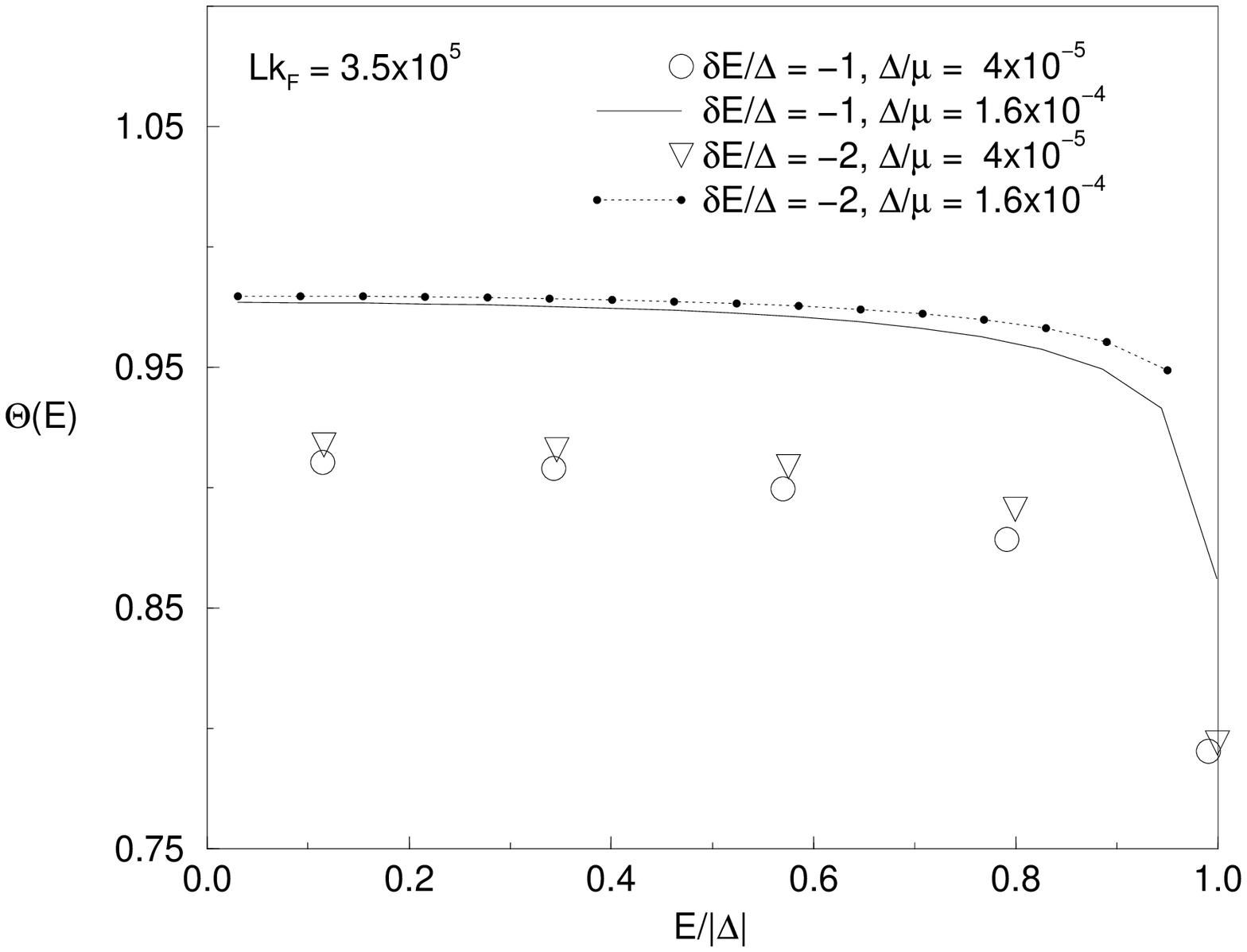,height=10.0cm}}
\caption[]{Localized Andreev states in the region II. Their
spacing is a smooth function of energy. Their energies hardly
depend on the position of the chemical potential with respect to
the maximum of the potential.}
\label{fm12L5}
\end{figure}

The results depicted in
FIG. \ref{fm12L5} correspond to $\delta E = - \Delta$ and $\delta E = -2 \Delta$.
So we have the space separation of Andreev states in the whole energy
interval. No interference occurs, and the spacing $\Theta$ is a smooth
function of  energy.  The
differences between  $\delta E = - \Delta$ and $\delta E = -2 \Delta$ are small.
For the smaller gap
value the number of states is of the same order as the degenerate
states for mode 6 in FIG. \ref{f10L1D2}. The Andreev levels hardly
depend on $\delta E$, since they never reach the middle of the 
constriction and do not experience the potential over there.



\section{Conclusions and prospects}
\label{conclusions}

In conclusion, we present here a theory of 
Andreev states for a shallow long adiabatic constriction.
The system differs from traditional models of S-N-S junctions
by the absence of electron-hole symmetry and exhibits
at least four disctict energy scales that determine the
interference of the components of Andreev states.
Such constrictions can be possibly realized in semiconductor
heterostructures. 

We have started with a completely semiclassical
approach to the states and reached in the first approximation
a good agreemement with exact solutions of Bogoliubov equations.
On the top of this, the exact Andreev levels may exhibit three
distinct types of oscillating behavoir. These oscillations
arise from interference of the transmitted and reflected 
waves in the middle of the constriction. 
We have extended the semiclassical theory to account for this interference
and were able to reach quantitative agreement with numerical results.

In this paper, we assumed no phase difference between the superconductors.
It would be interesting to calculate the Andreev states in
the presence of such phase difference and finally find the supercurrent
in the structure. It is already clear that the supercurrent
would exhibit an interesting dependence on controllable
system parameters, in particular, on $\delta E$. This makes
such calculation interesting in view of possible experiments.

\acknowledgments

This work is a part of the research programme of the "Stichting voor
Fundamenteel Onderzoek der Materie"~(FOM), and we acknowledge the financial
support from the "Nederlandse Organisatie voor Wetenschappelijk Onderzoek"
~(NWO). It is our pleasure to 
acknowledge many fruitful discussions with N. ChtChelkatchev.

\appendix
\section{Interference of Andreev states}
\label{quasiclasmod}
\begin{figure}[htb]
\centerline{\epsfig{figure=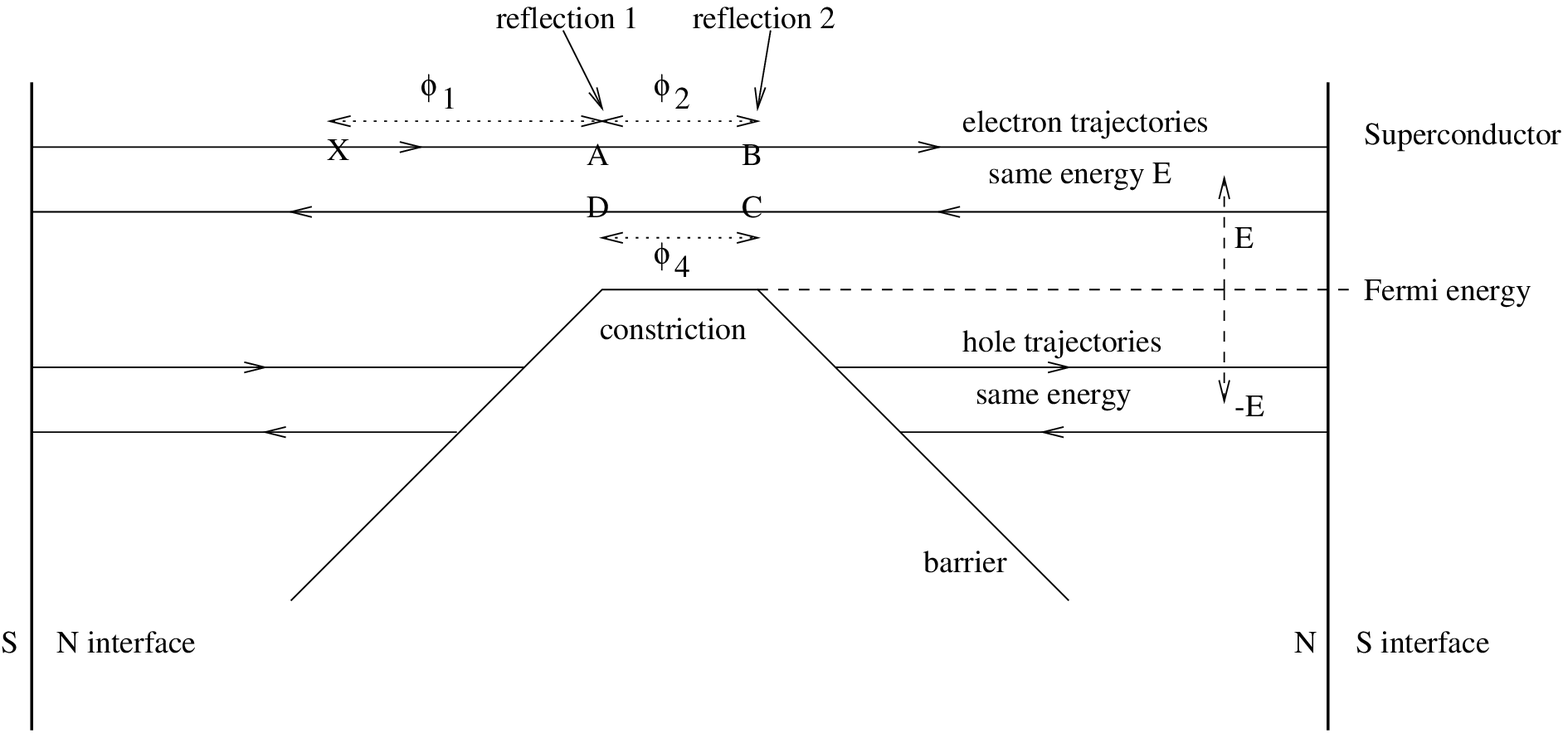,height=8.0cm}}
\caption[]{Schematical picture of our constriction, including
the paths of the electron and hole branches of the Andreev bound
states. The Fermi energy coincides with the flat part of
the potential at the constriction. In the quasiclassical
derivation reflections of the electron wave are accounted for which
are taken to occur at the the entrance and exit of the
constriction. Reflection 1 occurs at the point A, which coincides spatially
with point D, and reflection 2 occurs at point B (or C).}
\label{Dipmodel}
\end{figure}
The shape of the oscillations can be understood in quasiclassical
terms if the interference of the waves is taken into account. 
The configuration of the
system and the parameters used are shown in FIG. \ref{Dipmodel}.

To evaluate the quantization condition, we consider the amplitude 
of the wave that starts propagating from point X, goes all the way  
through the constriction and gets back to the same point.
The starting point can be chosen arbitrarily. 
Each part of the trajectory
contributes a factor of the form $e^{i\phi}$ to the amplitude. The
phases $\phi_1$, $\phi_2$ and $\phi_4$ are shown in the figure,
 $\phi_3$ is the propagation phase from B to C, and $\phi_5$
is the phase from D to X. If there is no reflection at A, B, C and D
\begin{equation}
\label{Ampnoref}
{\rm Amplitude} = e^{i\phi_1}e^{i\phi_2}e^{i\phi_3}e^{i\phi_4}e^{i\phi_5} = 1,
\end{equation}
in which the second equality is the condition for an Andreev
bound state. This quantization condition can be written as
\begin{equation} 
\label{ABSphi}
\phi_1 + \phi_2 + \phi_3 + \phi_4 + \phi_5 = 2 \pi n
\end{equation}
Using the symmetries
\begin{equation} 
\label{symphi}
\phi_1 + \phi_5 = \phi_3 \equiv 2 \phi_b
\hspace{4mm}{\rm and}\hspace{4mm}\phi_2 = \phi_4 \equiv  \phi_t
\end{equation}
one can write down the equality
\begin{equation} 
\label{delET}
2 \pi \hbar \frac{{\rm d}n}{{\rm d}E} = 2 T_t + 4 T_b,
\end{equation}
where $T_t$ is the time for the classical motion between
the points A and B, and $T_b$
is the time it takes to get from the interface to the constriction top. 
Here we made use
of the quasiclassical relation  between the phase and the time of the classical motion
$\hbar \frac{{\rm d}\phi}
{{\rm d}E} = T$.
Eq. (\ref{delET}) can be considered as a slight generalization
of Eq. (\ref{ETrelation2}).

Now we extend the analysis  by assuming
 the possibility of relections at the points A, B, C and D.
For each point four coefficients are to be defined, namely
$t_{LR}$ and $t_{RL}$ for transmission from left to right
and from right to left respectively, and $r_{LR}$ and $r_{RL}$,
similarly, for reflection. By summing up all possible ways of
propagation, from the point X to the right and back to that point coming
from the left, and accounting for all reflections and transmissions
one arrives at
\begin{equation} 
\label{Ampwref} 
{\rm Amplitude} = e^{i(\phi_1 + \phi_5)}\lbrack r_{LR}^{(1)} +
\frac{t_{LR}^{(1)}t_{RL}^{(1)}{\rm A}_{AD}}{1-r_{RL}^{(1)}
{\rm A}_{AD}}\rbrack = 1,
\end{equation}
in which the second equality is the quantization condition. The quantity
${\rm A}_{AD}$ is the amplitude to get from A to D and is given by
\begin{equation}  
\label{AmprefAD}  
{\rm A}_{AD}= e^{i(\phi_2 + \phi_4)}\lbrack r_{LR}^{(2)} +
\frac{t_{LR}^{(2)}t_{RL}^{(2)}e^{i\phi_3}}{1-r_{RL}^{(2)}e^{i\phi_3}}\rbrack. 
\end{equation}
Again using the phases $\phi_b$ and $\phi_t$, solving Eq. (\ref{Ampwref})
for ${\rm A}_{AD}$, and using Eq. (\ref{AmprefAD}),
one obtains the symmetric form
\begin{equation}  
\label{ABSref}
 e^{2i\phi_t}\hspace{2mm}\frac{z^{(1)}e^{i 2 \phi_b} + r_{RL}^{(1)}}
{1-e^{i 2 \phi_b}r_{LR}^{(1)}}\hspace{2mm}
\frac{z^{(2)}e^{i 2 \phi_b} + r_{LR}^{(2)}}{1-e^{ i 2\phi_N}r_{RL}^{(2)}}=1,
\end{equation}
in which $z^{(1)}$ and $z^{(2)}$ are defined by
\begin{equation}   
\label{z12}
z^{(1,2)}\equiv t_{LR}^{(1,2)}t_{RL}^{(1,2)}-r_{LR}^{(1,2)}r_{RL}^{(1,2)}.
\end{equation}
We now study the general expression (\ref{ABSref}) using
simplifying assumptions. If we disregard scattering phases and
their energy dependence we can write
\begin{equation}  
\label{rRL1}
r_{LR}^{(1)}=r_{LR}^{(2)}=r_{RL}^{(1)}=r_{LR}^{(2)}=\sqrt{R},
\end{equation}
$R$ being the reflection coefficient. By that we find, that
$z^{(1,2)}=-1$ and the quantization condition (\ref{ABSref}) becomes
\begin{equation}
\label{ABSrefR}
 e^{2i(\phi_t + 2 \phi_b)}{\left(\frac{1-e^{-i 2 \phi_b}\sqrt{R}}
{1-e^{i 2 \phi_b}\sqrt{R}}\right)}^2 = 1,
\end{equation}
which is equivalent to
\begin{equation}
\label{ABSphiref}
2(\phi_t + 2 \phi_b) + 4\hspace{1mm}{\rm atan}\left(\frac{\sin (2\phi_b)\sqrt{R}}
{1-\cos (2 \phi_b)\sqrt{R}}\right) = 2\pi n.
\end{equation}
For the energy derivative this gives
\begin{equation} 
\label{delETref}
2 \pi \hbar \frac{{\rm d}n}{{\rm d}E} = 2 T_t + 4 T_b +
4 T_b\frac{\cos(2 \phi_b)\sqrt{R} - R}{1+R-2\cos(2 \phi_b)\sqrt{R}}.
\end{equation}
We rewrite this using our definition of $\Theta$ and obtain the desired
expression, see Eq. (\ref{Theta_osc_0}),
\begin{equation}
\label{Theta_osc}
\Theta(E) = 1 +\frac {2 T_b}{T_t+2 T_b}\frac{\cos(2 \phi_b)\sqrt{R} - R}
{1+R-2\cos(2 \phi_b)\sqrt{R}}
\end{equation}
If there are no reflections, $\Theta$ is equal to 1.
If reflections are accounted for, Eq. (\ref{Theta_osc}) implies,
that for $R\ll 1$ small modulations below the value of 1
are found in plotting the quantity $\delta E(T_t + 2T_b)/\pi$
as a function of $E$ or $E/\Delta$.
If $R\simeq 1$ periodic dips below the value of
1 are coming out, completely in agreement of the structure of
FIG. \ref{f0L3}. For a constant $R$ the dips lie at equal height. In
reality $R$ is energy dependent at $E \simeq E_c$ and decreases with increasing
energy. Recalling the definition of $\phi_b,\phi_t$, we see
that the spacing is determined by the energy scale $E_{i2}$
(provided $E \simeq E_c$) and the oscillation period is
determined by the bigger energy scale $E_{i1}$.

Similar consideration can be performed for hole oscillations
in the region Ia. This results in oscillations with
a period determined by  difference of phases accumulated 
respectively by electrons and holes during their travel
though the middle of the constriction.

The irregular oscillations in the region Ib can be also
quantified along these lines. However, here the reflection coefficients
for the electron and hole wave are different, and different
phase differences are accumulated by electrons and holes
in the constriction middle and beyond. The typical quasiperiod
is determined by the phase $\phi_b$, that varies with energy
in a slowest way.

\end{document}